\documentclass{llncs}

\usepackage{latexsym}

\input{amssym.def}

\input{amssym}

\newbox\tempa
\newbox\tempb
\newdimen\tempc
\def\mud#1{\hfil $\displaystyle{\mathstrut #1}$\hfil}
\def\rig#1{\hfil $\displaystyle{#1}$}
\def\irulehelp#1#2#3{\setbox\tempa=\hbox{$\displaystyle{\mathstrut #2}$}%
		        \setbox\tempb=\vbox{\halign{##\cr
	\mud{#1}\cr
	\noalign{\vskip\the\lineskip}%
	\noalign{\hrule height 0pt}%
	\rig{\vbox to 0pt{\vss\hbox to 0pt{${\; #3}$\hss}\vss}}\cr
	\noalign{\hrule}%
	\noalign{\vskip\the\lineskip}%
	\mud{\copy\tempa}\cr}}%
		      \tempc=\wd\tempb
		      \advance\tempc by \wd\tempa
		      \divide\tempc by 2 }
\def\irule#1#2#3{{\irulehelp{#1}{#2}{#3}%
		     \hbox to \wd\tempa{\hss \box\tempb \hss}}}

\def\ex{\exists}
\def\fa{\forall}
\def\ra{\rightarrow}
\def\lra{\longrightarrow}

\begin{document}

\title{Automated theorem proving in first-order logic modulo:
on the difference between type theory
and set theory}

\author{Gilles Dowek
\institute{INRIA-Rocquencourt, B.P. 105, 78153 Le Chesnay Cedex,
France.
\texttt{Gilles.Dowek@inria.fr http://coq.inria.fr/\~{}dowek}}}

\date{}

\maketitle

\begin{abstract}
{\em Resolution modulo} is a first-order theorem proving method
that can be applied both to first-order presentations of simple type theory
(also called higher-order logic) and to set theory. When it is applied
to some first-order presentations of type theory, it simulates exactly
higher-order resolution. In this note, we compare how it
behaves on type theory and on set theory.
\end{abstract}

\bigskip

\section*{}

Higher-order theorem proving (e.g. higher-order resolution
\cite{Andrews71,Huet72,Huet73}) is different from first-order theorem
proving in 
several respects. First, the first-order unification algorithm has to
be replaced by the higher-order one \cite{Huet75,Huet76}. Even then, 
the resolution rule alone is not complete but another rule called {\em the
splitting rule} has to be added. At last, the skolemization rule is more
complicated \cite{Miller83,Miller87}.  

On the other hand, higher-order logic, also called simple type theory,
can be expressed as a first-order theory \cite{Davis}, and
first-order theorem proving methods, such as first-order resolution,
can be used for this theory. Of course, first-order resolution with
the axioms of this theory is much less efficient than higher-order
resolution. However, we can try to understand higher-order resolution
as a special automated theorem proving method designed for this theory. 
A motivation for this project is that it is very unlikely
that such a method applies only to this theory, but it
should also apply to similar theories such as extensions of type
theory with primitive recursion or set theory.

In \cite{modulo}, together with Th.~Hardin and C.~Kirchner, we have proposed a
theorem proving method for first-order logic, called {\em resolution
modulo}, that when applied to a first-order presentation of 
type theory simulates exactly higher-order resolution. Proving the
completeness of this method has required to introduce a new
presentation of first-order logic, called {\em deduction modulo} that
separates clearly computation steps and deduction steps.  

Resolution modulo can be applied both to type theory
and to set theory. The goal of this note is to
compare how resolution modulo works for one theory and the other.
In order to remain self contained, we will first
present shortly the ideas of deduction modulo and resolution modulo.

\bigskip

\section{Resolution modulo}

\subsection{Deduction modulo}

In deduction modulo, the notions of language, term and proposition are
that of (many sorted) first-order logic. But, a theory is formed with a
set of axioms $\Gamma$ {\em and a congruence 
$\equiv$} defined on propositions. In this paper, all congruences will
be defined by confluent rewrite systems (as
these rewrite systems are defined on propositions and propositions
contain binders, these rewrite systems are in fact {\em combinatory
reduction systems} \cite{KlopOostromRaamsdonk}).
Propositions are supposed to be identified modulo the congruence
$\equiv$. Hence, the deduction rules must take into account this
equivalence. For instance, the {\em modus ponens} cannot be stated as
usual
$$\irule{A \Rightarrow B~~~A}{B}{}$$
but, as the two occurrences of $A$ need not be identical, but need only
to be congruent, it must be stated
$$\irule{A' \Rightarrow B~~~A}{B}{\mbox{if $A \equiv A'$}}$$
In fact, as the congruence may identify implications with other
propositions, a slightly more general formulation is needed 
$$\irule{C~~~A}{B}{\mbox{if $C \equiv A \Rightarrow B$}}$$
All the rules of natural deduction or sequent calculus may be stated
in a similar way, see \cite{modulo,normalization} for more details.

As an example, in arithmetic, in natural deduction modulo, we can
prove that $4$ is an even number: 
$$\irule{\irule{\irule{}
                      {\fa x~x = x}
                      {\mbox{axiom}}
               }
               {2 \times 2 = 4}
               {\mbox{$(x,x = x,4)$ $\fa$-elim}}
        }
        {\ex x~2 \times x = 4}
        {\mbox{$(x,2 \times x = 4,2)$ $\ex$-intro}}$$
Substituting the variable $x$ by the term $2$ in the proposition 
$2 \times x = 4$ yields the proposition $2 \times 2 = 4$, 
that is congruent to $4 = 4$. The transformation of one proposition
into the other, that requires several proof steps in natural
deduction, is dropped from the proof in deduction modulo. It is just a
computation that need not be written, because everybody
can re-do it by him/herself. 

In this case, the congruence can be defined by a rewriting system
defined on terms 
$$0 + y \lra y$$
$$S(x) + y \lra S(x+y)$$
$$0 \times y \lra 0$$
$$S(x) \times y \lra x \times y + y$$

Notice that, in the proof above, we do not need the axioms of addition
and multiplication. Indeed, these axioms are now redundant: since the
terms $0 + y$ and $y$ are congruents, the axiom $\fa y~0 + y = y$ is
congruent to the equality axiom $\fa y~y = y$. Hence, it can be
dropped. In other words, this axiom has been built-in the congruence
\cite{Plotkin72,Andrews71,Stickel}. 

The originality of deduction modulo is that we have introduced the
possibility to define the congruence directly on propositions with
rules rewriting atomic propositions to arbitrary ones.
For instance, in the theory of integral rings, we can take the rule
$$x \times y = 0 \lra x = 0 \vee y = 0$$
that rewrites an atomic proposition to a disjunction.

Notice, at last, that deduction modulo is not a true extension of
first-order logic. Indeed, it is proved in \cite{modulo} that for every
congruence $\equiv$, we can find a theory ${\cal T}$ such that $\Gamma
\vdash P$ is provable modulo $\equiv$ if and only if ${\cal T} \Gamma
\vdash P$ is provable in ordinary first-order logic. Of course, the
provable propositions are the same, but the proofs are very different.

\subsection{Resolution modulo}

When the congruence on propositions is induced by a congruence on
terms, automated theorem proving can be performed like in first-order
logic, for instance with the resolution method, provided the
unification algorithm is replaced by an {\em equational unification}
algorithm modulo this congruence. 
Equational unification problems can be solved by the {\em narrowing}
method \cite{Fay,Hullot,JouannaudKirchner}.  
The method obtained this way, called {\em equational resolution}
\cite{Plotkin72,Stickel}, is complete.

The situation is different when the congruence identifies atomic
propositions with non atomic ones. For instance, in the theory of
integral rings, the proposition 
$$a \times a = 0 \Rightarrow a = 0$$ 
is provable because it reduces to 
$$(a = 0 \vee a = 0) \Rightarrow a = 0$$ 
Hence the proposition 
$$\ex y~(a \times a = y \Rightarrow a = y)$$ 
is also provable. But, with the clausal form of its negation
$$a \times a = Y$$
$$\neg a = Z$$
we cannot apply the resolution rule successfully, because the terms $a \times a$
and $a$ do not unify.

Hence, we need to introduce a new rule that detects that the literal
$a \times a = Y$ has an instance that is reducible by the rewrite rule 
$$x \times y = 0 \lra x = 0 \vee y = 0$$
instantiates it, reduces it and puts it in clausal form again. 
We get this way the clause
$$a = 0$$
that can be resolved with the clause $\neg a = Z$. 

Hence, the rewrite rules have to be divided into two sets: the set ${\cal
E}$ of rules rewriting terms to terms 
that are used by the equational unification algorithm and
the set of rule ${\cal R}$ rewriting atomic propositions to
arbitrary ones and that are used by this new rule called 
{\em extended narrowing}. The
system obtained this way is called {\em extended narrowing and
resolution} or simply {\em resolution modulo}. Figure \ref{ResNar}
gives a formulation of this method where unification problems are
postponed as constraints. 
\begin{figure}
\begin{center}
\noindent\framebox{\parbox{\textwidth}{
Extended resolution:
$$
\irule{\{A_{1}, \ldots, A_{n}, B_{1},\ldots, B_{m}\}/E_1
         ~~~
         \{\neg C_{1}, \ldots, \neg C_{p}, D_{1}, \ldots, D_{q}\}/E_2}
     {\{B_{1}, \ldots, B_{m}, D_{1}, \ldots, D_{q}\}/
      E_{1} \cup E_{2} \cup \{A_{1} =_{{\cal E}} \ldots
       A_{n} =_{{\cal E}} C_{1} =_{{\cal E}} \ldots
         C_{p}\}}
     {}$$
Extended narrowing:
$$\irule{C/E}
 	{\mbox{{\em cl}}(C[r]_p)/(E \cup \{C_{|p} =_{{\cal E}}l\})}
        {\mbox{if $l \lra r \in {\cal R}$}}
$$
\caption{Resolution modulo}
\label{ResNar}
}}
\end{center}
\end{figure}
A proposition is said to be provable with this
method when, from the clausal form of its negation, we can deduce an
empty clause constrained by a ${\cal E}$-unifiable set of equations.

Transforming axioms into rewrite rules enhances the efficiency of
automated theorem proving as shown by this very simple example.

\medskip
\noindent
{\em Example.}
To refute the theory
$P_{1} \Leftrightarrow (Q_{2} \vee P_{2})$,  
...,
$P_{i} \Leftrightarrow (Q_{i+1} \vee P_{i+1})$,  
...,
$P_{n} \Leftrightarrow (Q_{n+1} \vee P_{n+1})$,  
$P_{1}$, 
$Q_{2} \Leftrightarrow \bot$, 
...,
$Q_{n+1}\Leftrightarrow \bot$,
$P_{n+1} \Leftrightarrow \bot$, 
resolution yields $4n+2$ clauses 
$$\neg P_{1}, Q_{2}, P_{2}$$ 
$$\neg Q_{2}, P_{1}$$
$$\neg P_{2}, P_{1}$$
$$...$$
$$\neg P_{i}, Q_{i+1}, P_{i+1}$$ 
$$\neg Q_{i+1}, P_{i}$$
$$\neg P_{i+1}, P_{i}$$
$$...$$
$$\neg P_{n}, Q_{n+1}, P_{n+1}$$ 
$$\neg Q_{n+1}, P_{n+1}$$
$$\neg P_{n+1}, P_{n+1}$$
$$P_{1}$$
$$\neg Q_{2}$$
$$...$$
$$\neg Q_{n+1}$$
$$\neg P_{n+1}$$
While, in resolution modulo, the propositions
$P_{i} \Leftrightarrow (Q_{i+1} \vee P_{i+1})$,
$Q_{i} \Leftrightarrow \bot$
and 
$P_{n+1} \Leftrightarrow \bot$
can be transformed into rewrite rules
$$P_{i} \lra Q_{i+1} \vee P_{i+1}$$
$$Q_{i} \lra \bot$$
$$P_{n+1} \lra \bot$$
The only proposition left is 
$P_{1}$. It reduces to $\bot \vee  ... \vee \bot$ 
and its clausal form is hence the empty clause.

Of course, reducing the proposition $P_{1}$ has a cost, but this cost
is much lower than that of the non deterministic search of a
refutation resolution with the clauses above. Indeed, the reduction
process is deterministic because the rewrite system is confluent.

\subsection{Cut elimination and completeness}

Resolution modulo is not complete for all congruences. For
instance, take the congruence induced by the rewrite rule
$$A \lra A \Rightarrow B$$
The proposition $B$ has a proof in sequent calculus modulo

\hspace*{4.5cm}
$\irule{\irule{\irule{\irule{\irule{}{A \vdash A}{\mbox{axiom}}
                              ~~~~~~~~~~~~
                              \irule{\irule{}
                                           {B \vdash B}
                                           {\mbox{axiom}}
                                    }
                                    {A, B \vdash B}
                                    {\mbox{weak.-left}}}
                             {A, A \vdash B}
                             {\mbox{$\Rightarrow$-left}}
                      }
                      {A \vdash B}
                      {\mbox{contr.-left}}
               }
               {\vdash A}
               {\mbox{$\Rightarrow$-right}}
         ~~~~~~~~~~~~~~~~~~~~~~~~~~~~~~~~~~~~~~~~~
         \irule{\irule{\irule{}
                             {A \vdash A}
                             {\mbox{axiom}}
                       ~~~~~~~~~~~~
                       \irule{\irule{}
                                    {B \vdash B}
                                    {\mbox{axiom}}
                             }
                             {A, B \vdash B}
                             {\mbox{weak.-left}}}
                      {A, A \vdash B}
                      {\mbox{$\Rightarrow$-left}}
               }
               {A \vdash B}
               {\mbox{contr.-left}}
        }
        {\vdash B}
        {\mbox{cut}}$

\noindent but it is not provable by resolution modulo.
Indeed, the clausal form of the negation of the proposition $B$ is the
clause
$$\neg B$$
and neither the extended resolution rule nor the extended narrowing
rule can be applied successfully.

However, it may be noticed that the proposition $B$ has no cut free
proof in sequent calculus modulo. Hence sequent calculus modulo this
congruence does not have the cut elimination property. We have proved
in \cite{modulo} that resolution modulo is complete for all congruences
$\equiv$ such that the sequent calculus modulo $\equiv$ has the cut
elimination property. Together with B.~Werner, we have proved  in
\cite{normalization} that cut 
elimination holds modulo a large class of congruences and conjectured
that it holds modulo all congruences that can be defined
by a confluent and terminating rewrite system.

When cut elimination does not hold, only propositions that have a cut
free proof are proved by resolution modulo.

\bigskip

\section{Simple type theory and set theory}

\subsection{Simple type theory}

Simple type theory is a many-sorted first-order theory. 
The sorts of simple type theory, called {\em simple types}, are
defined inductively as follows. 
\begin{itemize}
\item $\iota$ and $o$ are simple types,
\item if $T$ and $U$ are simple types then $T \ra U$ is a simple type.
\end{itemize}

\smallskip
\noindent
As usual, we write $T_{1} \ra ... \ra T_{n} \ra U$ for the type $T_{1}
\ra ...~(T_{n} \ra U)$.

The language of simple type theory contains the individual symbols
\begin{itemize}
\item $S_{T,U,V}$ of sort $(T \ra U \ra V) \ra (T \ra U) \ra T \ra V$,
\item $K_{T,U}$ of sort $T \ra U \ra T$,
\item $\dot{\vee}$ of sort $o \ra o \ra o$,
\item $\dot{\neg}$ of sort $o \ra o$, 
\item $\dot{\fa}_{T}$ of sort $(T \ra o) \ra o$,
\end{itemize}
the function symbols
\begin{itemize}
\item $\alpha_{T,U}$ of rank $(T \ra U, T, U)$,
\end{itemize}
and the predicate symbol
\begin{itemize}
\item $\varepsilon$ of rank $(o)$.
\end{itemize}
As usual, we write $(t~u)$ for the term $\alpha(t,u)$ and 
$(t~u_{1}~...~u_{n})$ for $(...(t~u_{1})~...~u_{n})$. 

Usual presentations of simple-type theory \cite{Church40,Andrews86} define
propositions as terms of type $o$. 
But, as we want type theory to be a first-order theory, we introduce a
predicate symbol $\varepsilon$ that transforms a term of type $o$ into
a genuine proposition. Then, we need an axiom relating the proposition 
$\varepsilon(\alpha(\alpha(\dot{\vee},x),y))$ and the
proposition $\varepsilon(x) \vee \varepsilon(y)$. For instance,
the axiom 
$$\fa x~\fa y~(\varepsilon(\dot{\vee}~x~y)
\Leftrightarrow (\varepsilon(x) \vee \varepsilon(y)))$$
This axiom can be built in the congruence, if we take the rewrite rule 
$$\varepsilon(\dot{\vee}~x~y) \lra 
 \varepsilon(x) \vee \varepsilon(y)$$
This leads to the rewrite system of the figure \ref{HOL}. 
\begin{figure} \begin{center}
\noindent\framebox{\parbox{\textwidth}{{\small
$$(S~x~y~z) \lra (x~z~(y~z))$$
$$(K~x~y) \lra x$$
$$\varepsilon(\dot{\neg}~x) \lra \neg \varepsilon(x)$$
$$\varepsilon(\dot{\vee}~x~y) \lra
\varepsilon(x) \vee \varepsilon(y)$$ 
$$\varepsilon(\dot{\fa}~x) \lra \fa y~\varepsilon(x~y)$$
\caption{Rewriting rules for simple type theory}
\label{HOL}
}}}
\end{center}\end{figure}
This rewrite system is confluent because it is orthogonal and we prove in
\cite{tphols} that it is strongly normalizing. Hence, the congruence is
decidable. 

It is proved in \cite{normalization} that deduction modulo this congruence
has the cut elimination property, i.e. every proposition
provable in sequent calculus modulo this congruence has a cut free proof.

\subsection{Set theory}

The language of Zermelo's set theory is formed with the binary predicate
symbols $\in$ and $=$. This theory contains the axioms of
equality and the following axioms.\\  
pair:
$$\fa x~\fa y~\ex z~\fa w~(w \in z \Leftrightarrow (w = x \vee w =
y))$$
union:
$$\fa x~\ex y~\fa w~(w \in y \Leftrightarrow \ex z~(w \in z \wedge z
\in x))$$
power set:
$$\fa x~\ex y~\fa w~(w \in y \Leftrightarrow \fa z~(z \in w
\Rightarrow z \in x))$$
subset scheme:
$$\fa x_{1} ... \fa x_{n}~\fa y~\ex z~\fa w~(w \in z \Leftrightarrow
(w \in y \wedge P))$$
where $x_{1}, ..., x_{n}$ are the free variables of $P$ minus $w$.\\
To these axioms, we may add the extensionality axiom, the foundation
axiom, the axiom of infinity, the replacement scheme and the axiom of choice.

To have a language for the objects of the theory we may skolemize
these axioms introducing the function symbols $\{\}$, $\bigcup$,
${\cal P}$ and $f_{x_{1}, ..., x_{n}, w, P}$.
We then get the axioms 
$$\fa x~\fa y~\fa w~(w \in \{\}(x,y) \Leftrightarrow (w = x \vee w =
y))$$
$$\fa x~\fa w~(w \in \bigcup(x) \Leftrightarrow \ex z~(w \in z \wedge z
\in x))$$
$$\fa x~\fa w~(w \in {\cal P}(x) \Leftrightarrow \fa z~(z \in w
\Rightarrow z \in x))$$
$$\fa x_{1} ... \fa x_{n}~\fa y~\fa w~(w \in f_{x_{1}, ..., x_{n},
w, P}(x_{1}, ..., x_{n}, y)  \Leftrightarrow (w \in y \wedge P))$$
Then, these axioms may be built in the congruence with the rewrite system of 
figure \ref{Set}.
\begin{figure} \begin{center}
\noindent\framebox{\parbox{\textwidth
}{
{\small
$$w \in \{\}(x,y) \lra w = x \vee w = y$$
$$w \in \bigcup(x) \lra \ex z~(w \in z \wedge z \in x)$$
$$w \in {\cal P}(x) \lra  \fa z~(z \in w \Rightarrow z \in x)$$
$$v \in f_{x_{1}, ..., x_{n},w, P}(y_{1}, ..., y_{n}, z)  \lra v \in z
\wedge 
[y_{1}/x_{1}, ..., y_{n}/x_{n}, v/w]P$$
\caption{Rewriting rules for set theory}
 \label{Set}
}}}
\end{center}\end{figure}
This rewrite system is confluent because it is orthogonal. But it does
not terminate. A counter-example is 
M.~Crabb\'{e}'s proposition.
Let $C$ be the term $\{x \in a~|~\neg x \in x\}$ i.e. $f_{w, \neg w
\in w}(a)$. We have 
$$x \in C \lra x \in a \wedge \neg x \in x$$
Hence, writing $A$ for the proposition $C \in C$ and $B$ for the proposition 
$C \in a$ we have
$$A \lra B \wedge \neg A$$ 
This permits to construct the infinite reduction sequence
$$A \lra B \wedge \neg A \lra B \wedge \neg (B \wedge \neg A) \lra
...$$
Up to our knowledge, the decidability of this congruence is open.

Deduction modulo this congruence does not have the cut elimination
property. A counter example is again Crabb\'{e}'s proposition (see
\cite{Hallnas,Ekman} for a discussion). As we have seen, this
proposition $A$ rewrites to a proposition of the form $B \wedge \neg
A$. Hence, the proposition $\neg B$ has the following proof 

\hspace{4.5cm}
$\irule{\irule{\irule{\irule{}
                             {B \vdash B}
                             {\mbox{axiom}}
                       ~~~~~~~~~~~~
                       \irule{\irule{\irule{\irule{\irule{\irule{\irule{}
                                                                       {A \vdash A}
                                                                       {\mbox{axiom}}
                                                                }
                                                                {A, B \vdash A}
                                                                {\mbox{weakening-left}}
                                                         }
                                                         {A, B, \neg A \vdash}
                                                         {\mbox{$\neg$-left}}
                                                  }
                                                  {A, A \vdash}
                                                  {\mbox{$\wedge$-left}}
                                           }
                                           {A \vdash}
                                           {\mbox{contraction-left}}
                                    }
                                    {\vdash \neg A}
                                    {\mbox{$\neg$-right}}
                                 }
                                 {B \vdash \neg A}
                                 {\mbox{weakening-left}}
                      }
                      {B \vdash A}
                      {\mbox{$\wedge$-right}}
                ~~~~~~~~~~~~~~~~~~~~~~~~~~~~~~~~~~~~~~~~~~~
                \irule{\irule{\irule{\irule{\irule{}
                                                  {A \vdash A}
                                                  {\mbox{axiom}}}
                                           {A, B \vdash A}
                                           {\mbox{weakening-left}}
                                    }
                                    {A, B, \neg A \vdash}
                                    {\mbox{$\neg$-left}}
                             }
                             {A, A \vdash}
                             {\mbox{$\wedge$-left}}
                      }
                      {A \vdash}
                      {\mbox{contraction-left}}
               }
               {B \vdash}
               {\mbox{cut}}
        }
        {\vdash \neg B}
        {\mbox{$\neg$-right}}$

\noindent
but it is easy to check that the proposition $\neg B$, i.e. $\neg
f_{w, \neg w \in w}(a) \in a$, has no cut free proof.

\bigskip

\section{Resolution modulo in type theory and in set theory}

\subsection{Resolution modulo in type theory}

In the rewrite system of figure \ref{HOL}, the first two rules 
$$(S~x~y~z) \lra (x~z~(y~z))$$
$$(K~x~y) \lra x$$
rewrite terms to terms and are used by the unification algorithm. The 
three others
$$\varepsilon(\dot{\neg}~x) \lra \neg \varepsilon(x)$$
$$\varepsilon(\dot{\vee}~x~y) \lra
\varepsilon(x) \vee \varepsilon(y)$$ 
$$\varepsilon(\dot{\fa}~x) \lra \fa y~\varepsilon(x~y)$$
rewrite propositions to propositions and are used by the extended
narrowing rule. 

Equational unification modulo the rules $S$ and $K$ is related to
higher-order unification. Actually since the
reduction of combinators is slightly weaker than the reduction of
$\lambda$-calculus, unification modulo this reduction is slightly
weaker than higher-order unification \cite{DoughertyTCS93}. To have
genuine higher-order unification, we have to take another formulation
of type theory using explicit substitutions instead of
combinators (see section \ref{holsigma+}).

The extended narrowing modulo the rules $\dot{\neg}$, $\dot{\vee}$ and
$\dot{\fa}$ is exactly the splitting rule of higher-order
resolution. A normal literal unifies with the left member 
of such a rule if and only if its head symbol is a variable.

The skolemization rule in this language is related to the
skolemization rule of type theory. When we skolemize a
proposition of the form 
$$\fa x~\ex y~P$$
we introduce a function symbol $f$ of rank $(T,U)$ where $T$ is the
type of $x$ and $U$ the type of $y$ (not an individual symbol of type
$T \ra U$) and the axiom
$$\fa x~[f(x)/y]P$$
Hence, the Skolem symbol $f$ alone is not a term, but it permits to build a
term of type $U$ when we apply it to a term of type $T$. This is, in
essence, the higher-order skolemization rule, but formulated for the
language of combinators and not for $\lambda$-calculus. Again, we have
the genuine higher-order skolemization rule if we use the formulation
of type theory using explicit substitutions instead of
combinators (see section \ref{holsigma+}).

\subsection{Resolution modulo in set theory}

In set theory, there is no rule rewriting terms to terms. Hence,
unification in set theory is simply first-order unification. 
Converselly, all the rules of figure \ref{Set} rewrite propositions to
propositions and thus the extended narrowing is performed modulo all
theses rules.

In set theory, resolution modulo is incomplete. We have
seen that the proposition  
$$\neg f_{w, \neg w \in w}(a) \in a$$
has a proof in set theory, but it cannot be proved by the resolution
modulo method.
Indeed, from the clausal form of its negation 
$$f_{w, \neg w \in w}(a) \in a$$
we can apply neither the resolution rule nor the extended
narrowing rule successfully. 

\section{On the differences between set theory and type theory}

\subsection{Termination}

The first difference between resolution modulo in type theory and in set
theory is that the rewrite system is terminating in type theory and
hence all propositions have a normal form, while some propositions,
e.g. Crabb\'{e}'s proposition, have no normal form in set theory. 

Hence, during proof search, we can normalize all the clauses 
while this is impossible in set theory. Formally, the method 
modified this way requires a completeness proof. 

\subsection{Completeness}

Another difference is that, as type theory verifies the cut
elimination property, resolution modulo this congruence is complete,
while it is incomplete modulo the congruence of set theory. 

A solution to recover completeness may be to use an automated theorem
proving method
that searches for proofs containing cuts. For instance if we add a
rule allowing to refute the set of clauses $S$ by refuting both the
set $S \cup \{\neg P\}$ and the set $\{P\}$ then we can refute the proposition
$B$ above. 

Another direction is to search for another presentation of set theory
or for a restriction of this theory that enjoys termination and 
cut elimination. We conjecture that if we restrict the
subset scheme to {\em stratifiable} propositions in the sense of
W.V.O. Quine \cite{Quine},  we get a restriction of set theory that is
sufficient to express most mathematics, that terminates and that
verifies the cut elimination property. The cut elimination and
completeness results obtained by S.C.~Bailin \cite{Bailin88a,Bailin88b}
for his formulation of set theory let this conjecture be plausible. 

\subsection{Typing literals}

A minor difference is that when we try to prove a theorem of the
form ``for all natural numbers $x$, $P(x)$'', we have to formalize this
theorem by the proposition 
$$\fa x~(x \in {\Bbb N} \Rightarrow P(x))$$
in set theory. In contrast, in type theory, we can choose to take
$\iota$ for the type of natural numbers and state the theorem 
$$\fa x~P(x)$$
During the search, in set theory, extra literals of the form $x \in
{\Bbb N}$ appear and have to be resolved. 

\subsection{The role of unification and extended narrowing}

In resolution modulo, like in most other methods, the main difficulty
is to construct the terms that have to be substituted to the
variables.
In resolution modulo, these terms are constructed by two processes:
the unification algorithm and the extended narrowing rule.

The main difference between resolution modulo in type theory and in
set theory is the division of work between the unification
and the extended narrowing. In type theory, unification is quite powerful
and the extended narrowing is rarely used. In contrast, in set theory,
unification is simply first-order unification and all the work
is done by the extended narrowing rule. 

This difference reflects a deep difference on how mathematics are
formalized in a theory and the other. Indeed, the unification in type
theory is rich because there are rules that rewrite terms to terms and
these rules are there because the notion of function is primitive in
type theory.
When we have a function $f$ and an object $a$ we can
form the term $(f~a)$ and start rewriting this term to a normal
form. In set theory, there is no such term and a term alone can never
be reduced. 
Instead of forming the term $(f~a)$ we can
form a proposition expressing that $b$ is the image of $a$ by the
function $f$, $<a,b> \in f$, that then can be rewritten.

For example, in the proof of Cantor's theorem we have a function $f$
from a set $B$ to its power set and we want to form Cantor's set of objects
that do not belong to their image.

If $x$ is an element of $B$, in type theory we can express its image
$(f~x)$, then the term of type $o$ reflecting the proposition expressing
that $x$ belongs to  its image $(f~x~x)$, the term of type $o$
reflecting its negation $\dot{\neg} (f~x~x)$ and then Cantor's set 
$\lambda x~\dot{\neg} (f~x~x)$ that, with combinators, is expressed 
by the term 
$$C = (S~(K~\dot{\neg})~(S~(S~(K~f)~(S~K~K))~(S~K~K)))$$

In contrast, in set theory, we cannot form a term expressing the image
of $x$ by the function $f$. Instead of saying that $x$ does not belong
to its image we have to say that it does not belong to any objet that
happens to be its image.
$$C = \{x \in B~|~\fa y~(<x,y> \in f \Rightarrow \neg x \in y)\}$$
This requires to introduce two more logical symbols $\Rightarrow$ and
$\fa$. These symbols cannot be generated by the unification algorithm 
and are generated by the extended narrowing rule.

It is not completely clear what is the best division of work
between unification and extended narrowing. Experiences with type
theory show that the unification algorithm is usually well controlled
while the splitting rule is very productive. Loading the unification
and unloading the extended narrowing seems to improve efficiency.

However, two remarks moderate this point of view. First, in type
theory, the functions
that can be expressed by a term are very few. For instance, if we take
the type $\iota$ for the natural numbers and introduce two symbols $O$
and $Succ$ for zero and the successor function,
we can only express by a term the constant functions and the
functions adding a constant to their argument. The other functions are
usually expressed with the description operator (or the choice operator)
and hence as relations. We may enrich the language of combinators and 
the rewrite system, for instance with primitive recursion, but then it
is not obvious that unification is still so well controlled.

Another remark is that having a decidable and unitary unification
(such as first-order unification) permits to solve unification
problems on the fly instead of keeping them as constraints. This
permits to restrict the use of the extended narrowing rule.
For instance, in type theory, when we have a literal
$\varepsilon(P~x)$ and we apply the extended narrowing rule yielding
two literals $\varepsilon(A)$ and $\varepsilon(B)$ and a constraint 
$$\varepsilon(P~x) = \varepsilon(A \dot{\vee} B)$$
we keep this constraint frozen and we may need to apply the extended
narrowing rule to other literals starting with the variable $P$. 
In contrast, in set theory, if we have a literal $x \in P$ and we apply
the extended narrowing rule yielding two literals $y = a$ and $y = b$ and
a constraint $(x \in P) = (y \in \{a,b\})$. The substitution
$\{a,b\}/P$ can be immediately propagated to all the occurrences of $P$ 
initiating reductions that let the extended narrowing steps be useless.

\bigskip

\section{Advanced formulations of type theory and set theory}
\label{holsigma+}

As an illustration of this discussion, we want to compare
resolution modulo proofs of Cantor's theorem in 
type theory and in set theory. 
However, the presentations of type theory and set theory above
are a little too rough to be really practicable. In both cases, we shall
use a more sophisticated presentation where the language contains a
full binding operator. 

Indeed, in type theory, we want to express Cantor's set by the term
$$C = \lambda x~\dot{\neg}(f~x~x)$$
and not by the term
$$C = (S~(K~\dot{\neg})~(S~(S~(K~f)~(S~K~K))~(S~K~K)))$$
Similarily, in set theory we want to express this set as 
$$C = \{x \in B~|~\fa y~(<x,y> \in R \Rightarrow \neg x \in y)\}$$
where $<x,y>$ is a notation for the set $\{\{x,y\},\{x\}\}$ i.e.
$\{\}(\{\}(x,y),\{\}(x,x))$, and not by the term

\medskip
\noindent
$C = \{x \in B~|~\fa y~(
\fa u~((\fa v~(v \in u \Leftrightarrow (\fa w~(w \in v$

\hfill
$\Leftrightarrow (w = x \vee w = y))
\vee \fa w~(w \in v \Leftrightarrow w = x)))) \Rightarrow u \in R)
\Rightarrow \neg x \in y)\}$\footnote{In the presentation of set
theory above, there is no instance of the
subset scheme for the proposition 
$$\fa y~(<x,y> \in R \Rightarrow \neg x \in y)$$
because it contains Skolem symbols. Hence, we 
replace the proposition $<x,y> \in R$ by the equivalent one 
{\scriptsize $$\fa u~((\fa v~(v \in u \Leftrightarrow (\fa w~(w \in v
\Leftrightarrow (w = x \vee w = y))
\vee \fa w~(w \in v \Leftrightarrow w = x)))) \Rightarrow u \in R)$$}
Then we can build the set $C$ with the function symbol introduced by
the skolemization of this instance of the scheme. 
The proposition $x \in C$ is then provably equivalent to 
$x \in B \wedge \fa y~(<x,y> \in R \Rightarrow \neg x \in y)$ but it
does not reduce to it.}

\medskip

For type theory, such a first-order presentation with a general binding
operator has been proposed in \cite{holsigma}. It uses an expression of
$\lambda$-calculus as a first-order language based on de Bruijn indices
and explicit substitutions. In this presentation, the sorts are of the form 
$\Gamma \vdash T$ or $\Gamma \vdash \Delta$ where $T$ is a simple type
and $\Gamma$ and $\Delta$ are finite sequences of simple types.
The language contains the following symbols 

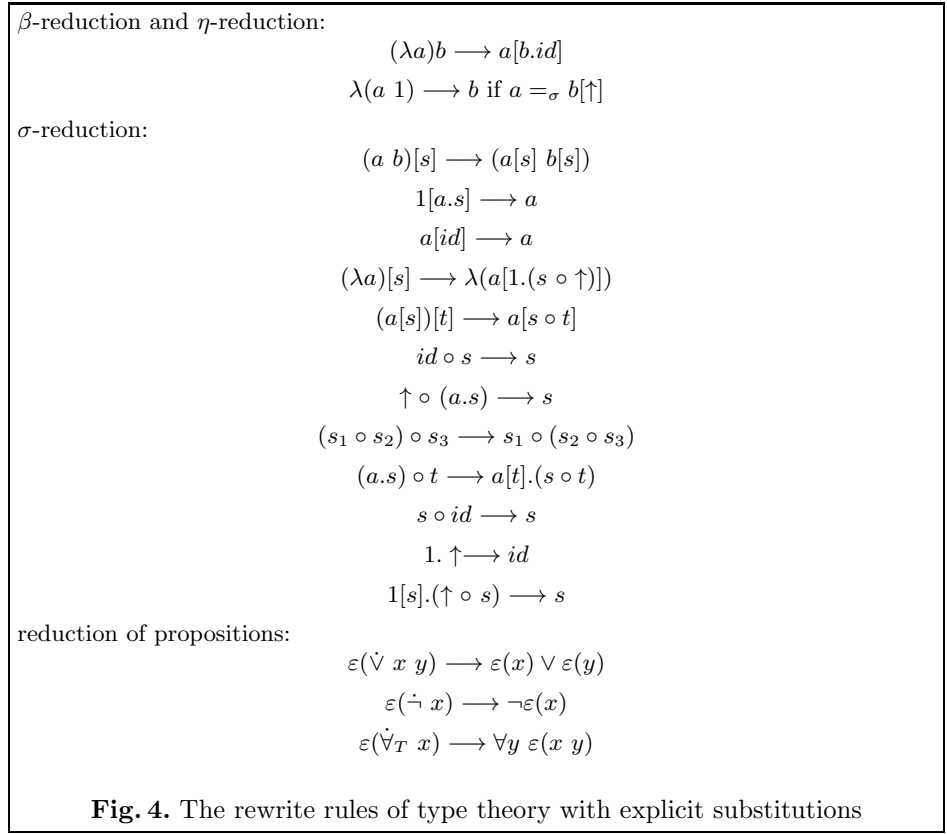
\begin{figure} \begin{center}
\noindent\framebox{\parbox{\textwidth}{
{\small
$\beta$-reduction and $\eta$-reduction:
$$(\lambda a) b \lra a[b.id]$$
$$\lambda (a~1) \lra b~\mbox{if $a =_{\sigma} b[\uparrow]$}$$

$\sigma$-reduction:
$$(a~b)[s] \lra (a[s]~b[s])$$
$$1[a.s] \lra a$$
$$a[id] \lra  a$$
$$(\lambda a)[s] \lra  \lambda (a[1.(s ~\circ \uparrow)])$$
$$(a[s])[t] \lra  a[s \circ t]$$
$$id \circ s \lra  s$$
$$\uparrow \circ ~ (a.s) \lra  s$$
$$(s_1 \circ s_2) \circ s_3 \lra  s_1 \circ (s_2 \circ s_3)$$
$$(a.s) \circ t \lra  a[t].(s \circ t)$$
$$s \circ id \lra  s$$
$$1. \uparrow \lra  id$$
$$1[s].(\uparrow \circ ~ s) \lra s$$

reduction of propositions:
$$\varepsilon(\dot{\vee}~x~y) \lra \varepsilon(x) \vee \varepsilon(y)$$
$$\varepsilon(\dot{\neg}~x) \lra \neg \varepsilon(x)$$
$$\varepsilon(\dot{\fa}_{T}~x) \lra \fa y~\varepsilon(x~y)$$
\caption{The rewrite rules of type theory with explicit substitutions}
\label{holsigma}
}}}
\end{center}\end{figure}

\begin{itemize}
\item $1_A^{\Gamma}$ of sort $A \Gamma \vdash A$,
\item $\alpha_{A,B}^{\Gamma}$ of rank $(\Gamma \vdash A \ra B, \Gamma
       \vdash A, \Gamma \vdash B)$,
\item $\lambda_{A,B}^{\Gamma}$ of rank $(A \Gamma \vdash B, \Gamma
\vdash A \ra B)$, 
\item $\mbox{[]}_{A}^{\Gamma,\Gamma'}$ of rank $\Gamma' \vdash A, \Gamma
\vdash \Gamma', \Gamma \vdash A)$,
\item $id^{\Gamma}$ of sort $\Gamma\vdash\Gamma$,
\item $\uparrow_{A}^{\Gamma}$ of sort $A\Gamma \vdash \Gamma$, 
\item $._{A}^{\Gamma,\Gamma'}$ of rank $(\Gamma \vdash A, \Gamma
\vdash \Gamma', \Gamma \vdash A \Gamma')$,
\item $\circ^{\Gamma,\Gamma',\Gamma''}$ of rank $(\Gamma \vdash
\Gamma'', \Gamma'' \vdash \Gamma', \Gamma \vdash \Gamma')$,
\item $\dot{\vee}$ of sort $\vdash o \ra o \ra o$,
\item $\dot{\neg}$ of sort $\vdash o \ra o$,
\item $\dot{\fa}_{T}$ of sort $\vdash (T \ra o) \ra o$,
\item $\varepsilon$ of rank $(\vdash o)$.
\end{itemize}
And the rewrite system is that of figure \ref{holsigma}.

A formulation of set theory with a general binder has been given in 
\cite{combinators}. But it is not expressed in a first-order setting
yet. Waiting for such a theory, for the example of Cantor's theorem, 
we add a constant $C$ and an {\em ad hoc} rewrite rule  
$$x \in C \lra x \in B \wedge \fa y~(<x,y> \in R \Rightarrow \neg x \in y)$$

\bigskip

\section{Three proofs of Cantor's theorem}

We now give three resolution modulo proofs 
of Cantor's theorem that there is no surjection from
a set to its power set. 
The first is in type theory with a function
expressing the potential surjection from a set to its power set. 
The second is also in type theory, but this potential surjection
is expressed by a relation. The last one is in set theory and the
surjection is, of course, expressed by a relation.

Automated theorem proving for Cantor's theorem
in type theory is discussed in  \cite{Huet72,Huet73,AMCP}.

\subsection{In type theory with a function}

In type theory, a set is expressed by a term of type $T \ra o$. Here, we
choose to consider only the set of all objects of type $\iota$. 
Its power set is the set of all objects of type $\iota \ra o$. Hence we
want to prove that there is no surjection from the type $\iota$ to
$\iota \ra o$. The first solution is to represent this potential
surjection by a function $f$ of type $\iota \ra \iota \ra o$. 
The surjectivity of this function can be expressed by the existence of
a right-inverse $g$ to this function, 
i.e. a function of type $(\iota \ra o) \ra \iota$ 
such that for all $x$, $(f~(g~x)) = x$. Using Leibniz' definition of
equality this proposition is written
$$\fa x~\fa p~(\varepsilon(p~(f~(g~x))) \Leftrightarrow \varepsilon(p~x))$$
Putting this proposition in clausal form yields the clauses
$$\neg \varepsilon(P~(f~(g~X))), \varepsilon(P~X)$$
$$\varepsilon(Q~(f~(g~Y))), \neg \varepsilon(Q~Y)$$
The search is described on figure 
\ref{fCantortypefonction}.
\begin{figure} \begin{center}
\noindent\framebox{\parbox{\textwidth}{
{\small
$$\begin{array}{rll}
1 &  & \neg \varepsilon(P~(f~(g~X))), \varepsilon(P~X)\\
2 &  & \varepsilon(Q~(f~(g~Y))), \neg \varepsilon(Q~Y)\\
3 & \mbox{narr.}~(1) & 
\neg \varepsilon(P~(f~(g~X))), \neg \varepsilon(R)/c_{1}\\
4 & \mbox{narr.}~(2) &
\varepsilon(Q~(f~(g~Y))), \varepsilon(S) /c_{2}\\
5 & \mbox{res.}~(3,4) &
\Box/c_{1},c_{2},c_{3},c_{4},c_{5}\\
\end{array}$$
with$$\begin{array}{ll}
c_{1} & (P~X) = \dot{\neg} R\\
c_{2} & (Q~Y) = \dot{\neg} S\\
c_{3} & (P~(f~(g~X))) = R\\
c_{4} & (Q~(f~(g~Y))) = S\\
c_{5} & (P~(f~(g~X))) = (Q~(f~(g~Y)))\\
\end{array}$$
\caption{Cantor's theorem in type theory with a function}
\label{fCantortypefonction}
}}}

\end{center}\end{figure}
It returns the empty clause constrained by the equations
$$\begin{array}{l}
(P~X) = \dot{\neg} R\\
(Q~Y) = \dot{\neg} S\\
(P~(f~(g~X))) = R\\
(Q~(f~(g~Y))) = S\\
(P~(f~(g~X))) = (Q~(f~(g~Y)))\\
\end{array}$$
that have the solution
$$\begin{array}{l}
X = Y = \lambda \dot{\neg}[\uparrow] (f[\uparrow]~1~1)\\
P = Q = \lambda (1~(g[\uparrow]~\lambda \dot{\neg}[\uparrow^{2}] (f[\uparrow^{2}]~1~1)))\\
R = S = (f~(g~\lambda \dot{\neg}[\uparrow]
(f[\uparrow]~1~1))~(g~\lambda \dot{\neg}[\uparrow](f[\uparrow]~1~1)))
\end{array}$$

\subsection{In type theory with a relation}

Instead of using the primitive notion of function of set theory, we
can code the functions as functional relations $R$ of type $\iota \ra
(\iota \ra o) \ra o$. The surjectivity and functionality of this
relation are expressed by the propositions 
$$E:\fa y~\ex x~\varepsilon(R~x~y)$$
and 
$$F:\fa x~\fa y~\fa z~(\varepsilon(R~x~y) \Rightarrow \varepsilon(R~x~z)
\Rightarrow \fa p~(\varepsilon(p~y) \Leftrightarrow \varepsilon(p~z)))$$
Putting these propositions in clausal form yields the clauses
$$\varepsilon(R~g(U)~U)$$
$$\neg \varepsilon(R~X~Y), \neg \varepsilon(R~X~Z), \neg
\varepsilon(P~Y), \varepsilon(P~Z)$$
$$\neg \varepsilon(R~X~Y), \neg \varepsilon(R~X~Z), \neg
\varepsilon(P~Z), \varepsilon(P~Y)$$

The search is then described on figure \ref{fCantortyperel}
where we simplify the constraints and substitute the solved
constraints at each step. 
\begin{figure} \begin{center}
\noindent\framebox{\parbox{\textwidth}{
{\small
$$\begin{array}{rll}
1 & & \varepsilon(R~g(U)~U)  \\
2 & & \neg \varepsilon(R~X~Y), \neg \varepsilon(R~X~Z), \neg
\varepsilon(P~Y), \varepsilon(P~Z)  \\
3 & & \neg \varepsilon(R~X~Y), \neg \varepsilon(R~X~Z), \neg
\varepsilon(P~Z), \varepsilon(P~Y)  \\
4 & \mbox{res.}~(1,2) &
\neg \varepsilon(R~g(Y)~Z), \neg \varepsilon(P~Y), 
\varepsilon(P~Z) \\
5 & \mbox{res.}~(1,4) &
\neg \varepsilon(P~Y), \varepsilon(P~Y) \\
6 & \mbox{five narr.}~(5) &
\neg \varepsilon(A_{1}), \neg \varepsilon(B_{1}),
\varepsilon(P~Y)/c_{1},c_{2} \\ 
7 & \mbox{res.}~(6,1) &
\neg \varepsilon(B_{1}),\varepsilon(P~Y)/c_{1},c'_{2} \\
8 & \mbox{four narr.}~(7) &
\varepsilon(A_{2}), \varepsilon(P~Y)/c_{1},c''_{2},c_{3} \\
9 & & \varepsilon(B_{2}), \varepsilon(P~Y)/c_{1},c''_{2},c_{3}\\
10 & \mbox{renaming}~(6) & 
\neg \varepsilon(A'_{1}), \neg \varepsilon (B'_{1}),
\varepsilon(P~Y)/ c_{1},c_{4} \\
11 & \mbox{res.}~(10,8) &
\neg \varepsilon(B'_{1}),
\varepsilon(P~Y)
/c_{1},c''_{2}, c_{3}, c'_{4} \\
12 & \mbox{res.}~(11,9) &
\varepsilon(P~Y)/
c_{1},c''_{2}, c_{3}, c''_{4} \\
13 & \mbox{five narr.}~(12) & 
\varepsilon(A_{3})/c_{1},c''_{2}, c_{3},c''_{4}, c_{5},c_{6} \\
14 & & 
\varepsilon(B_{3})/
c_{1},c''_{2}, c_{3}, c''_{4},c_{5},c_{6} \\
15 & \mbox{renaming}~(4) &
\neg \varepsilon(R~g(Y')~Z'), \neg \varepsilon(P~Y'), 
\varepsilon(P~Z') \\
16 & \mbox{res.}~(15,13) &
\neg \varepsilon(P~Y'), \varepsilon(P~Z')/ 
c_{1},c''_{2}, c_{3},c''_{4}, c_{5},c'_{6} \\
17 & \mbox{narr.}~(16) & 
\neg \varepsilon(P~Y'), \neg \varepsilon(Q)/ 
c_{1},c''_{2}, c_{3},c''_{4}, c_{5},c'_{6},c_{7}\\
18 & \mbox{res.}~(17,14) &
\neg \varepsilon(P~Y')/ 
c_{1},c''_{2}, c_{3},c''_{4},c_{5},c'_{6},c'_{7} \\
19 & \mbox{five narr.}~(18) &
\neg \varepsilon(A_{4}), \neg \varepsilon(B_{4})/
c_{1},c''_{2},
c_{3},c''_{4},c_{5},c'_{6},c'_{7},c_{8},c_{9} \\
20 & 
\mbox{res.}~(19,1) &
\neg \varepsilon(B_{4})/
c_{1},c''_{2},
c_{3},c''_{4},c_{5},c'_{6},c'_{7},c_{8},c'_{9}\\
21 & \mbox{four narr.}~(20) &
\varepsilon(A_{5})/
c_{1},c''_{2},
c_{3},c''_{4},c_{5},c'_{6},c'_{7},c_{8},c''_{9},c_{10}\\
22 & & \varepsilon(B_{5})/
c_{1},c''_{2},
c_{3},c''_{4},c_{5},c'_{6},c'_{7},c_{8},c''_{9},c_{10} \\
23 & \mbox{renaming}~(19) & 
\neg \varepsilon(A'_{4}), \neg \varepsilon(B'_{4})/
c_{1},c''_{2}, c_{3},c''_{4},c_{5},c'_{6},c'_{7},c_{8},c_{11} \\
24 & \mbox{res.}~(23,21) &
\neg \varepsilon(B'_{4})/
c_{1},c''_{2},
c_{3},c''_{4},c_{5},c'_{6},c'_{7},c_{8},c''_{9},c_{10},c'_{11}\\
25 & \mbox{res.}~(24,22) & 
\Box/
c_{1},c''_{2},c_{3},c''_{4},c_{5},c'_{6},c'_{7},c_{8},c''_{9},c_{10},c''_{11}
\\
\end{array}$$
with 
$$\begin{array}{ll}
c_{1} & (P~Y) = \dot{\neg} \dot{\fa} G_{1}\\
c_{2} & (G_{1}~W_{1}) = \dot{\neg} A_{1} \dot{\vee} \dot{\neg} B_{1}\\
c'_{2} & (G_{1}~W_{1}) = \dot{\neg} (R~g(U_{1})~U_{1})\dot{\vee} \dot{\neg} B_{1}\\
c''_{2} & (G_{1}~W_{1}) = \dot{\neg} (R~g(U_{1})~U_{1})\dot{\vee}
\dot{\neg} \dot{\fa} G_{2}\\ 
c_{3} & (G_{2}~y_{1}) = \dot{\neg} A_{2} \dot{\vee} \dot{\neg} B_{2}\\
c_{4} & (G_{1}~W'_{1}) = \dot{\neg} A'_{1} \dot{\vee} \dot{\neg} B'_{1}\\
c'_{4} & (G_{1}~W'_{1}) = \dot{\neg} A_{2} \dot{\vee} \dot{\neg} B'_{1}\\
c''_{4} & (G_{1}~W'_{1}) = \dot{\neg} A_{2} \dot{\vee} \dot{\neg} B_{2}\\
c_{5} & (P~Y) = \dot{\neg} \dot{\fa} G_{3}\\
c_{6} & (G_{3}~y_{2}) = \dot{\neg} A_{3} \dot{\vee} \dot{\neg} B_{3}\\
c'_{6} & (G_{3}~y_{2}) = \dot{\neg} (R~g(Y')~Z') \dot{\vee} \dot{\neg} B_{3}\\
c_{7} & (P~Z') = \dot{\neg} Q\\
c'_{7} & (P~Z') = \dot{\neg} B_{3}\\
c_{8} & (P~Y') = \dot{\neg} \dot{\fa} G_{4}\\
c_{9} &  (G_{4}~W_{2}) = \dot{\neg} A_{4} \dot{\vee} \dot{\neg} B_{4}\\
c'_{9} & (G_{4}~W_{2}) = \dot{\neg} (R~g(U_{2})~U_{2}) \dot{\vee} \dot{\neg} B_{4}\\
c''_{9} & (G_{4}~W_{2}) = \dot{\neg} (R~g(U_{2})~U_{2}) \dot{\vee}
\dot{\neg} \dot{\fa} G_{5}\\
c_{10} & (G_{5}~y_{3}) = \dot{\neg} A_{5} \dot{\vee} \dot{\neg} B_{5}\\
c_{11} &  (G_{4}~W'_{2}) = \dot{\neg} A'_{4} \dot{\vee} \dot{\neg} B'_{4}\\
c'_{11} &  (G_{4}~W'_{2}) = \dot{\neg} A_{5} \dot{\vee} \dot{\neg} B'_{4}\\
c''_{11} &  (G_{4}~W'_{2}) = \dot{\neg} A_{5} \dot{\vee} \dot{\neg} B_{5}\\
\end{array}$$
\caption{Cantor's theorem in type theory with a relation}
\label{fCantortyperel}
}}}
\end{center}\end{figure}
It returns the empty clause constrained by the equations 
$$\begin{array}{l}
(P~Y) = \dot{\neg} \dot{\fa} G_{1}\\
(G_{1}~W_{1}) = \dot{\neg} (R~g(U_{1})~U_{1})\dot{\vee}
\dot{\neg} \dot{\fa} G_{2}\\ 
(G_{2}~y_{1}) = \dot{\neg} A_{2} \dot{\vee} \dot{\neg} B_{2}\\
(G_{1}~W'_{1}) = \dot{\neg} A_{2} \dot{\vee} \dot{\neg} B_{2}\\
(P~Y) = \dot{\neg} \dot{\fa} G_{3}\\
(G_{3}~y_{2}) = \dot{\neg} (R~g(Y')~Z')
\dot{\vee} \dot{\neg} B_{3}\\
(P~Z') = \dot{\neg} B_{3}\\
(P~Y') = \dot{\neg} \dot{\fa} G_{4}\\
(G_{4}~W_{2}) = \dot{\neg} (R~g(U_{2})~U_{2}) \dot{\vee}
\dot{\neg} \dot{\fa} G_{5}\\
(G_{5}~y_{3}) = \dot{\neg} A_{5} \dot{\vee} \dot{\neg} B_{5}\\
(G_{4}~W'_{2}) = \dot{\neg} A_{5} \dot{\vee} \dot{\neg} B_{5}\\
\end{array}$$
that have the solution
$$\begin{array}{l}
P = \lambda \dot{\neg}[\uparrow](1~(g(C)[\uparrow]))\\
G_{1} = G_{2} = G_{3} = G_{4} = G_{5} = \lambda (\dot{\neg}[\uparrow]
(R[\uparrow]~g(C)[\uparrow]~1) \dot{\vee}[\uparrow] \dot{\neg}[\uparrow] (1~g(C)[\uparrow]))\\
Y = W_{1} = U_{1} = Y' = W_{2} = U_{2} = C\\
W'_{1} = y_{1}\\
Z' = y_{2}\\
W'_{2} = y_{3}\\
A_{2} = (R~g(C)~y_{1})\\
B_{2} = (y_{1}~g(C))\\
B_{3} = (y_{2}~g(C))\\
A_{5} = (R~g(C)~y_{3})\\
B_{5} = (y_{3}~g(C))\\
\end{array}$$
where $C = \lambda x~\dot{\fa} \lambda y~(\dot{\neg} (R~x~y) \dot{\vee}
\dot{\neg} (y~x))$,\\ i.e. 
$\lambda~\dot{\fa}[\uparrow] \lambda~(\dot{\neg}[\uparrow^2] 
(R[\uparrow^2]~2~1) \dot{\vee}[\uparrow^2]  \dot{\neg}[\uparrow^2](1~2))$.

\subsection{In set theory}

We consider a set $B$ and a potential surjection from this set to its
power set. We express this potential surjection by a set $R$. The
surjectivity and 
functionality of this set are expressed by the propositions
$$E:\fa y~(y \in {\cal P}(B) \Rightarrow \ex x~(x \in B \wedge <x,y> \in R))$$
$$F:\fa x~\fa y~\fa z~(<x,y> \in R \Rightarrow <x,z> \in R \Rightarrow y = z)$$
We use also the axiom of equality
$$L:\fa z~\fa x~\fa y~(x = y \Rightarrow \neg z \in x \Rightarrow \neg z \in
y)$$
The proposition $E$ reduces to the proposition
$$\fa u~(\fa y~(y \in u \Rightarrow y \in B)) 
\Rightarrow \ex x~(x \in B \wedge <x,u> \in R))$$
Putting this proposition in clausal form yields the clauses
$$y(U) \in U, <g(U),U> \in R$$
$$\neg y(U) \in B, <g(U),U> \in R$$
$$y(U) \in U, g(U) \in B$$
$$\neg y(U) \in B, g(U) \in B$$
The two other propositions yield the clauses
$$\neg <X,Y> \in R, \neg <X,Z> \in R, Y = Z$$
$$\neg X = Y, Z \in X, \neg Z \in Y$$

The search is described on figure \ref{fCantorens} where we 
simplify the constraints and substitute the solved constraints at each
step. Propagating the solved constraints may lead to new reductions
that require to put the proposition in clausal form again. This
explains that some resolution steps yield several clauses.
\begin{figure} \begin{center}
\noindent\framebox
{\parbox{\textwidth}
{{\small
$$\begin{array}{rll}
1 & & y(U) \in U, <g(U),U> \in R\\
2 & & \neg y(U) \in B, <g(U),U> \in R\\
3 & & y(U) \in U, g(U) \in B\\
4 & & \neg y(U) \in B, g(U) \in B\\
5 & & \neg <X,Y> \in R, \neg <X,Z> \in R, Y = Z\\
6 & & \neg X = Y, Z \in X, \neg Z \in Y\\
7 & \mbox{narr.}~(1) & 
y(C) \in B, <g(C),C> \in R\\
8 & & 
\neg <y(C),W> \in R,
\neg y(C) \in W,
<g(C),C> \in R\\
9 & \mbox{res.}~(7,2) & <g(C),C> \in R\\
10 & \mbox{narr.}~(3) &
y(C) \in B, g(C) \in B\\
11 &  & 
\neg <y(C),W> \in R,
\neg y(C) \in W,
g(C) \in B\\
12 & \mbox{res.}~(10,4) & g(C) \in B\\
13 & \mbox{res.}~ (9,5) &
 \neg <g(C),Z> \in R, C = Z\\
14 & \mbox{res.}~ (9,13) & C = C\\
15 & \mbox{res.} (14,6)&
Z \in B, \neg Z \in B, <Z,y_{1}(Z)> \in R\\
16 & & Z \in B, \neg Z \in B, Z \in y_{1}(Z)\\
17 & & 
\neg <Z,W_{1}> \in R, \neg Z \in W_{1}, \neg Z \in B, <Z,y_{1}(Z)> \in R\\
18 & & \neg <Z,W_{1}> \in R, \neg Z \in W_{1}, \neg Z \in B, Z \in y_{1}(Z)\\
19 & \mbox{res.}~(17,9) & 
\neg g(C) \in B,
  <g(C),y_{2}> \in R,
  <g(C),y_{1}(g(C))> \in R\\
20 & & \neg g(C) \in B,
  g(C) \in y_{2},
  <g(C),y_{1}(g(C))> \in R\\
21 & \mbox{res.}~(19,12) & 
<g(C),y_{2}> \in R,
  <g(C),y_{1}(g(C))> \in R\\
22 & \mbox{res.}~(20,12) & 
g(C) \in y_{2},
  <g(C),y_{1}(g(C))> \in R\\
23 & \mbox{res.}~ (18,9) & 
\neg g(C) \in B,
  <g(C),y_{3}> \in R, 
  g(C) \in y_{1}(g(C))\\
24 & & \neg g(C) \in B,
  g(C) \in y_{3},
  g(C) \in y_{1}(g(C))\\
25 & \mbox{res.}~ (23,12) & 
<g(C),y_{3}> \in R, 
  g(C) \in y_{1}(g(C))\\
26 & \mbox{res.}~ (24,12) & 
g(C) \in y_{3},
  g(C) \in y_{1}(g(C))\\
27 & \mbox{res.}~ (17,21) &
\neg g(C) \in y_{2}, 
\neg g(C) \in B,
<g(C),y_{1}(g(C))> \in R\\
28 & \mbox{res.}~(27,12) &
\neg g(C) \in y_{2}, 
<g(C),y_{1}(g(C))> \in R\\
29 & \mbox{res.}~ (28,22) &
<g(C),y_{1}(g(C))> \in R\\
30 & \mbox{res.}~(18,25) &
\neg g(C) \in y_{3}, 
\neg g(C) \in B, 
g(C) \in y_{1}(g(C))\\
31 & \mbox{res.}~(30,12) &
\neg g(C) \in y_{3}, 
     g(C) \in y_{1}(g(C))\\
32 & \mbox{res.}~ (31,26) &
  g(C) \in y_{1}(g(C))\\
33 & \mbox{res.}~(29,13) & 
C = y_{1}(g(C))\\
34 & \mbox{res.}~ (33,6) & 
Z \in B, \neg Z \in y_{1}(g(C))\\
35 & & 
\neg <Z,W_{2}> \in R, \neg Z \in W_{2}, \neg Z \in y_{1}(g(C))\\
36 & \mbox{res.}~ (35,32) & 
\neg <g(C),W_{2}> \in R, \neg g(C) \in W_{2}\\
37 & \mbox{res.}~ (36,9) & 
\neg g(C) \in B,
  <g(C),y_{4}> \in R\\
38 & & \neg g(C) \in B,
  g(C) \in y_{4}\\
39 & \mbox{res.}~ (37,12) &
<g(C),y_{4}> \in R\\
40 & \mbox{res.}~ (38,12) &
g(C) \in y_{4}\\
41 & \mbox{res.}~ (36,39) & 
\neg g(C) \in y_{4}\\
42 & \mbox{res.}~ (41,40) & 
\Box\\
\end{array}$$
\caption{Cantor's theorem in set theory}  
\label{fCantorens}
}}}
\end{center}\end{figure}
It returns the empty clause.

\subsection{Remarks}

The termination and completeness issues are not addressed by these
examples because, even in set theory, Cantor's theorem has a cut free proof
and the search involves only terminating propositions.

The proof in set theory is longer because several steps
are dedicated to the treatment of typing literals that are repeatedly
resolved with the clause $(12)$. 

In type theory with a function, only two extended narrowing steps are needed
to generate the symbol $\dot{\neg}$ in the term $\lambda
~\dot{\neg}[\uparrow](f[\uparrow]~1~1)$ (i.e. $\lambda x~\dot{\neg}(f~x~x)$)
that expresses Cantor's set. 
In type theory with a relation, four extended narrowing steps are
needed to generate the term 
$\lambda~\dot{\fa}[\uparrow] \lambda~(\dot{\neg}[\uparrow^2] 
(R[\uparrow^2]~2~1) \dot{\vee}[\uparrow^2]  \dot{\neg}[\uparrow^2](1~2))$
(i.e. $\lambda x~\dot{\fa} \lambda y~(\dot{\neg} (R~x~y) \dot{\vee}
\dot{\neg} (y~x))$) that expresses Cantor's set.
The term expressing Cantor set is thus mostly constructed by the
unification algorithm in the first case and mostly constructed by the
extended narrowing rule in the second.
In set theory, like in type theory with a relation, the term
expressing Cantor's set is mostly constructed by the extended narrowing rule.

In this case, a single step is needed because we have taken the {\em ad
hoc} rule 
$$x \in C \lra x \in B \wedge \fa y~(<x,y> \in R \Rightarrow \neg x
\in y)$$
But in a reasonable formulation of set theory several steps would be needed.

Notice, at last, that in the proof in type theory with a relation, the
term expressing Cantor's set is constructed several times, because the
constraints are frozen while in set theory, because the constraints
are solved on the fly, this term is constructed only twice and
propagated. To avoid this redundancy in type theory with a relation, 
it would be a good idea to solve as soon as possible the
constraints $c_{1}$ and $c_{2}$.

\bigskip
 
\section*{Conclusion}

Using a single automated theorem proving method for type theory and
for set theory permits a comparison. 

Although the use of a typed (many-sorted) language can be criticized, type
theory has several advantages for automated theorem proving: typing
permits to avoid typing literals, it enjoys termination and cut
elimination, and the possibility to form a term $(f~a)$ expressing the
image of an object by a function avoids indirect definitions.

This motivates the search of a type-free formalization of
mathematics, that also enjoys termination and cut elimination and where 
functions are primitive.


\begin{thebibliography}{99.}

\bibitem{Andrews71}
P.B. Andrews.
Resolution in type theory.
{\em The Journal of Symbolic Logic}, 36, 3 (1971), pp. 414-432.
 
\bibitem{Andrews86}
P.B.~Andrews, 
An introduction to mathematical logic and type theory: to 
truth through proof, {\em Academic Press} (1986).

\bibitem{AMCP}
P.B.~Andrews, D.A.~Miller, E.~Longini Cohen, and F.~Pfenning,
Automating higher-order logic,
W.W.~Bledsoe and D.W.~Loveland (Eds.), 
{\em Automated theorem proving: after 25 years},
Contemporary Mathematics Series 29, 
American Mathematical Society (1984), pp.~169-192.

\bibitem{Bailin88a}
S.C.~Bailin, 
A normalization theorem for set theory, 
{\em The Journal of Symbolic Logic}, 53, 3 (1988), pp.~673-695.

\bibitem{Bailin88b}
S.C.~Bailin, A $\lambda$-unifiability test for set theory,
{\em Journal of Automated Reasoning}, 4 (1988), pp.~269-286.

\bibitem{Church40}
A.~Church, 
A formulation of the simple theory of types, 
{\em The Journal of Symbolic Logic}, 5 (1940), pp.~56-68.

\bibitem{Davis}
M.~Davis, 
Invited commentary to \cite{Robinson68},
A.J.H. Morrell (Ed.) 
{\em Proceedings of the International Federation for Information 
Processing Congress}, 1968, North Holland (1969) pp. 67-68.

\bibitem{DoughertyTCS93}
D.J.~Dougherty,
Higher-order unification via combinators,
{\em Theoretical Computer Science}, 114 (1993), pp.~273-298.

\bibitem{combinators}
G.~Dowek, 
Lambda-calculus, combinators and the comprehension scheme,
M.~Dezani-Ciancaglini and G.~Plotkin (Eds.),
{\em Typed Lambda Calculi and Applications},
Lecture notes in computer science 902, Springer-Verlag (1995), pp.~154-170.
{\em Rapport de Recherche} 2565, INRIA (1995).

\bibitem{tphols}
G.~Dowek,
Proof normalization for a first-order formulation of higher-order
logic, 
E.L.~Gunter and A.~Felty (Eds.), 
{\em Theorem Proving in Higher-order Logics},
Lecture notes in computer science 1275, Springer-Verlag (1997), pp.~105-119.
{\em Rapport de Recherche} 3383, INRIA (1998).

\bibitem{modulo}
G.~Dowek, Th.~Hardin, and C.~Kirchner, 
Theorem proving modulo, 
{\em Rapport de Recherche} 3400, INRIA (1998).

\bibitem{holsigma}
G.~Dowek, Th.~Hardin, and C.~Kirchner,
HOL-$\lambda\sigma$: an intentional first-order 
expression of higher-order logic, to appear in {\em Rewriting
Techniques and Applications} (1999). {\em Rapport de Recherche} 3556,
INRIA (1998).  

\bibitem{normalization}
G.~Dowek and B.~Werner, 
Proof normalization modulo, 
{\em Rapport de Recherche} 3542, INRIA (1998).

\bibitem{Ekman}
J.~Ekman, 
Normal proofs in set theory, {\em Doctoral thesis}, Chalmers
University of Technology and University of G\"oteborg (1994).

\bibitem{Fay}
M.~Fay,
First-order unification in an equational theory,
{\em Fourth Workshop on Automated Deduction} (1979), pp.~161-167.

\bibitem{Hallnas}
L.~Halln\"as, On normalization of proofs in set theory, {\em Doctoral
thesis}, University of Stockholm (1983).

\bibitem{Huet72}
G.~Huet, 
Constrained resolution: a complete method for higher order logic,
{\em Ph.D.}, Case Western Reserve University (1972).

\bibitem{Huet73}
G.~Huet, 
A mechanization of type theory,
{\em International Joint Conference on Artificial Intelligence}
(1973), pp.~139-146.

\bibitem{Huet75}
G.~Huet, 
A unification algorithm for typed lambda calculus,
{\em Theoretical Computer Science},
1,1 (1975), pp.~27--57.

\bibitem{Huet76}
G.~Huet,
R\'{e}solution d'\'{e}quations dans les Langages d'Ordre 1,2, ...,
$\omega$,
{\em Th\`{e}se d'\'{E}tat}, Universit\'{e} de Paris VII (1976).

\bibitem{Hullot}
J.-M.~Hullot,
Canonical forms and unification,
W.~Bibel and R.~Kowalski (Eds.)
{\em Conference on Automated Deduction},
Lecture Notes in Computer Science 87,
Springer-Verlag (1980),
pp.~318-334.

\bibitem{JouannaudKirchner}
J.-P.~Jouannaud and C.~Kirchner,
Solving equations in abstract algebras: a rule-based survey of
unification, 
J.-L.~Lassez and G.~Plotkin (Eds.)
{\em Computational logic. Essays in honor of Alan Robinson},
MIT press (1991), pp.~257--321.

\bibitem{KlopOostromRaamsdonk}
J.W.~Klop, V.~van Oostrom, and F.~van Raamsdonk, 
Combinatory reduction systems:
introduction and survey, {\em Theoretical Computer Science}, 121
(1993), pp.~279-308.

\bibitem{Miller83}
D.A.~Miller, 
Proofs in higher order logic, {\em Ph.D.}, 
Carnegie Mellon University (1983).

\bibitem{Miller87}
D.A.~Miller, 
A compact representation of proofs, {\em Studia Logica}, 46, 4
(1987).

\bibitem{Plotkin72}
G.~Plotkin,
Building-in equational theories,
{\em Machine Intelligence},
7 (1972), pp.~73-90.

\bibitem{Quine}
W.V.O.~Quine, 
Set theory and its logic, {\em Belknap press} (1969).

\bibitem{Robinson68}
J.A. Robinson.
New directions in mechanical theorem proving.
A.J.H. Morrell (Ed.) 
{\em Proceedings of the International Federation for Information 
Processing Congress}, 1968, North Holland (1969), pp. 63-67.

\bibitem{Robinson70}
J.A. Robinson.
A note on mechanizing higher order logic.
{\em Machine Intelligence} 5, Edinburgh
university press (1970), pp. 123-133. 

\bibitem{Stickel}
M.~Stickel,
Automated deduction by theory resolution,
{\em Journal of Automated Reasoning},
4, 1 (1985), pp.~285-289.

\end{thebibliography}
\end{document}